\documentclass[11pt]{article}
\usepackage{fullpage}
\usepackage{graphicx} 
\usepackage{amsmath, amssymb, amsthm}
\usepackage{bbm}
\usepackage{parskip} 
\usepackage[ruled,noend]{algorithm2e}

\SetAlFnt{\small}
\SetAlCapFnt{\small}
\SetAlCapNameFnt{\small}
\SetAlCapHSkip{0pt}
\IncMargin{-\parindent}

\usepackage{booktabs}
\usepackage{multirow}
\usepackage{thm-restate}
\usepackage{xcolor}
\usepackage[colorlinks=true, citecolor=blue]{hyperref}
\usepackage{natbib}
\usepackage{enumitem}
\usepackage{url}
\usepackage{array}
\usepackage{comment}
\usepackage{todonotes}
\usepackage{color-edits}
\addauthor{cp}{red}

\title{Incentive-Aware Machine Learning; Robustness, Fairness, Improvement \& Causality\footnote{This literature review was recently published in SIGEcom Exchanges.}}

\author{
    Chara Podimata\\ MIT \\ \href{podimata@mit.edu}{\texttt{podimata@mit.edu}}
}

\date{}

\theoremstyle{plain}
\newtheorem{theorem}{Theorem}[section]

\theoremstyle{definition}

\theoremstyle{remark}
\newtheorem{remark}[theorem]{Remark}

\usepackage{mdframed}

\newcommand{\calX}{\mathcal{X}}
\newcommand{\calY}{\mathcal{Y}}

\newcommand{\hx}{\hat{x}}

\newcommand{\bbR}{\mathbb{R}}
\newcommand{\calD}{\mathcal{D}}
\newcommand{\fst}{f^{\star}}

\newcommand{\eps}{\varepsilon}
\newcommand{\calF}{\mathcal{F}}
\newcommand{\fopt}{f^{\textsf{OPT}}}
\newcommand{\sign}{\textsf{sign}}
\newcommand{\Reg}{\text{Reg}}
\newcommand{\1}{\mathbbm{1}}
\newcommand{\val}{\textsf{val}}
\newcommand{\cost}{\textsf{cost}}

\newcommand{\E}{\mathbb{E}}
\newcommand{\hst}{h^{\star}}
\newcommand{\calH}{\mathcal{H}}

\begin{document}

\maketitle

\begin{abstract}
The article explores the emerging domain of incentive-aware machine learning (ML), which focuses on algorithmic decision-making in contexts where individuals can strategically modify their inputs to influence outcomes. It categorizes the research into three perspectives: \emph{robustness}, aiming to design models resilient to ``gaming''; \emph{fairness}, analyzing the societal impacts of such systems; and \emph{improvement/causality}, recognizing situations where strategic actions lead to genuine personal or societal improvement. The paper introduces a unified framework encapsulating models for these perspectives, including offline, online, and causal settings, and highlights key challenges such as differentiating between gaming and improvement and addressing heterogeneity among agents. By synthesizing findings from diverse works, we outline theoretical advancements and practical solutions for robust, fair, and causally-informed incentive-aware ML systems. 
\end{abstract}

\section{Introduction}

Machine Learning (ML) algorithms are deeply embedded in various aspects of modern life, influencing everything from enhancing daily conveniences and shaping online purchasing behavior to making critical decisions in areas such as hiring, loan approvals, college admissions, and probation rulings. Given the high stakes of these decisions, individuals often have strong incentives to strategically modify the data they provide to these algorithms to secure more favorable outcomes. For instance, individuals might open additional credit accounts or take other steps to improve their credit scores before applying for a loan. In the context of college admissions, applicants may retake standardized tests like the GRE, enroll in test preparation courses, or even switch schools to boost their class rankings, all in efforts to present themselves as more competitive candidates.

Such instances of ``strategic adaptation'' have been extensively documented across disciplines including Economics, CS, and Public Policy~\cite{bjorkegren2020manipulation,dee2019causes,dranove2003more,greenstone2022can,gonzalez2019slippery,chang2024s}. The challenge arises when decision-makers deploying ML algorithms fail to account for these adaptations, potentially undermining the original goals of the policies the algorithms are intended to support. For example, in college admissions, a student's decision to change schools solely to improve their class ranking may not necessarily reflect a substantive improvement in their qualifications.

It is important to note that not all strategic adaptations are inherently problematic. Some represent attempts to ``game'' the system (e.g., switching schools for a better ranking), while others involve genuine efforts at self-improvement (e.g., dedicating more time to study). The distinction between these types of adaptations underscores the nuanced nature of this phenomenon and its implications for algorithmic decision-making.

\emph{What should decision-makers do when individuals are incentivized to alter the data they provide to ML algorithms in pursuit of better outcomes? And even if the learner manages to robustify (or calibrate) their algorithms to account for such behavior, what are the societal implications?} These are some of the central questions addressed by the emerging field of \emph{incentive-aware ML} (also known as ``strategic classification'' or ``performative prediction'').\footnote{Throughout this article, we use the terms ``incentive-aware'' and ``strategic'' ML interchangeably. The author prefers ``incentive-aware'' as it more comprehensively captures the considerations arising from agents' behaviors. However, ``strategic'' is more commonly used in the literature.}

The purpose of this article is to provide an introduction to incentive-aware ML and an overview of the key results in the field. We categorize the literature on incentive-aware ML into three main perspectives: \emph{robustness}, \emph{fairness}, and \emph{improvement \& causality}. While some papers contain elements of multiple categories, we classify them based on their primary focus or central contribution. Broadly speaking, the ``robustness'' perspective adopts the viewpoint of the decision-maker, assuming that agents always attempt to ``game'' the decision rule. The goal in this context is to design algorithms that achieve optimality despite strategic adaptations by the agents. The ``fairness'' perspective examines the downstream societal impacts of algorithmic decision-making under varying assumptions about the agents' capacity to strategically adapt. Lastly, the ``improvement \& causality'' perspective recognizes that not all strategic adaptations are harmful; in some cases, agents' adaptations in response to decision-making algorithms lead to genuine, fundamental improvements rather than merely fooling the algorithm. The distinctions among these perspectives, as well as the models and settings considered, will be formalized in the following section.

This article is organized as follows: Section~\ref{sec:model} formalizes all the different formulations of the incentive-aware ML problem while giving some example reference papers for each modeling assumption; Section~\ref{sec:robustness} presents a breakdown of the main contributions from the literature from the \emph{robustness} perspective; Section~\ref{sec:causality-improvement} outlines the results that have been obtained from the \emph{improvement \& causality perspective}; Section~\ref{sec:fairness} discusses the \emph{fairness} perspective. Finally, Section~\ref{sec:conclusion} offers some parting thoughts on where the literature stands and where we should go next (according to the author's personal opinions, at least).

\section{Overview of Models}\label{sec:model}

In the problem of incentive-aware learning, there is an interaction between a \emph{principal} (aka \emph{learner}, \emph{decision-maker}) and \emph{agents}. The problem has been studied both in the \emph{offline} (i.e., where there is one decision that is made by the principal and then the interaction stops) and \emph{online} setting (i.e., where there are sequential decisions). Before we outline each setting, let us introduce some common notation. 

Let $\calX \subseteq \bbR^d$ the \emph{feature} space and $\calY = \{0, 1\}$ (resp. $\calY \subseteq [0,1]$ for linear regression) the \emph{label} (resp. response) space. We assume that the label (resp. response) is $y = \hst(x)$, where $\hst: \calX \to \calY$ is called the \emph{ground truth} function (which is not necessarily linear).\footnote{Some works on incentive-aware linear regression assume that $y = \hst(x) + \eps$ where $\eps$ is some small, zero mean noise, but that will not constitute a material difference in this article.} We will denote by $\calH$ the \emph{concept class} where $\hst$ belongs to.
 
Let $\ell: \calY \times \calY \to [0,1]$ be the principal's loss function. Different applications of interest (within the general incentive-aware learning literature) call for different loss functions for the principal. Examples of frequently used loss functions for classification tasks include: 
    \begin{enumerate}[label=(\roman*)]
        \item $0-1$ loss (e.g., \cite{chen2020learning}): $\ell( y, y') := \1 \left\{ \sign (y \cdot y') = 1\right\}$.\footnote{We use $\sign(x) = 1$ to denote that $x$ is positive and $\sign(x) = - 1$ otherwise.} 
        \item logistic loss (e.g., \cite{dong2018strategic}): $\ell( y, y' ) := \log \left( 1 + e^{-y \cdot y'} \right)$
        \item hinge loss (e.g., \cite{dong2018strategic}): $\ell( y, y' ) := \max \left\{ 0, 1 - y \cdot y'\right\} $
    \end{enumerate}
    For the regression tasks, the most commonly used loss is some $L_p$ norm.
    
    As is the case in traditional ML, the choice of loss function for the principal affects what algorithms should be used, and what guarantees can be obtained.

\subsection*{Offline Setting}

In the offline setting (e.g., \cite{hardt2016strategic}), we assume that the agents' features are drawn from some distribution $\calD$. The interaction between the principal and the agent can be viewed as a \emph{Stackelberg game} that plays out as follows: 

\vspace{10pt}
\begin{mdframed}
\begin{enumerate}
    \item Nature draws $x \sim \calD$.
    \item The principal ---without knowing $x$--- commits to (and publicly announces) a decision-making rule $f: \calX \to \calY$.
    \item The agents observe $f$ and their point $(x,y)$.
    \item Given $f, x, y$, the agents choose $\hx(f)$ where $\hx(f; x,y) \in \calX$ is the best-response of the agent (given pair $(x,y)$) to the principal's rule $f$. 
    \item The agent reports point $(\hx(f; x, y), y)$ to the principal.
\end{enumerate}
\end{mdframed}
\vspace{10pt}

In Step (4), we are using $\hx(f; x, y)$ abstractly; we are going to specify how it is computed later on. At a high level, $\hx(f; x,y)$ is such that the agent obtains a better standing with regards to $f$ (e.g., the agent gets classified as $+1$ from $f$ in classification settings); see ``Agents' Response'' for details. To simplify notation, we write $\hx(f)$ (instead of $\hx(f; x, y)$ when clear from context. 

For now, let's assume that when the agents best respond to a decision-making rule, they are \emph{merely} trying to ``game'' it. We will contrast this approach to the Causality viewpoint, highlighted below.

In the ``robustness'' perspective, the principal's goal is to find a function $\fst \in \calF$ (where $\calF: \calX \to \calY$ is a hypothesis class over which we are searching) such that: 
\begin{equation}\label{eq:accuracy-offline}
\fst = \arg \min_{f \in \calF} \quad \E_{x \sim \calD} \left[ \ell \left(\hst(x),  f \left( \hx(f)\right)\right)\right]  
\end{equation}
In words, in the ``robustness'' perspective for the offline learning setting, the principal's goal is to find a function that minimizes the expected loss between the ground truth label (resp. response variable for regression) and the predicted label (resp. score) that function $f$ assigns to the (potentially) altered datapoint $\hx(f)$.

\subsection*{Online Setting}

In the online setting (e.g., \cite{dong2018strategic,chen2020learning,ahmadi2021strategic}, the interaction between the principal and the agents happens repeatedly over $T$ rounds. For every round $t \in [T]$, the interaction protocol is the following: 

\vspace{10pt}
\begin{mdframed}
\begin{enumerate}
    \item Nature chooses $x_t \in \calX$.
    \item The principal (without observing $x_t$) commits to (and publicly announces) decision-making rule $f_t \in \calF$. 
    \item The agent observes $f_t$ and their point $(x_t, y_t)$.
    \item The agent chooses $\hx_t(f_t; x_t, y_t)$ such that $\hx_t(f_t; x_t, y_t)$ is the agent's best response (given pair $(x_t, y_t)$) to rule $f_t$.
    \item The agent reports point $(\hx_t(f_t; x_t, y_t), y_t)$ to the principal.
\end{enumerate}
\end{mdframed}

\vspace{10pt}

As we did for the offline case, for ease of notation, we will simply write $\hx_t(f_t)$ in place of $\hx_t(f_t; x_t, y_t)$ whenever clear from context. In the online setting, we assume that the principal knows the agents' utility function, but not the original point, $x_t$. The choice of $\hx_t(f_t)$ in Step (4) depends on the agent's utility function; see ``Agent's Response'' below.

A couple of remarks are in order. First, the sequence $\{x_t\}_{t \in [T]}$ that the nature chooses can be \emph{adversarial}. Second, $\calF$ can be a general class of functions. That said, the current literature only focuses on \emph{linear} functions. Third, for the robustness perspective in online learning settings, we again assume that when the agent strategically adapts to a rule $f_t$, they can not influence their $y_t$ (i.e., $y_t$ remains the same both for $x_t$ and for the misreport $\hx_t$).

When we adopt the robustness perspective, the principal's goal is to minimize \emph{Stackelberg} regret defined as follows: 
\begin{equation}\label{eq:Stackelberg-regret}
\Reg(T) := \sum_{t \in [T]} \ell\left( \hst(x_t),f_t(\hx_t(f_t)) \right) - \min_{\fopt \in \calF} \sum_{t \in [T]} \ell \left( \hst(x_t) , \fopt\left(\hx_t\left( \fopt \right)\right)\right)
\end{equation}
Note that similar to the offline model, we are comparing the algorithm's performance to the best fixed rule $\fopt$ \emph{had you given the agents the opportunity to best respond}. In other words, we are comparing to the Stackelberg equilibrium rule.

\subsection*{Causality}

So far, in both the offline and online settings, we have assumed that even after the agent strategically adapts, their $y_t$ remains the same as it was prior to the adaptation (e.g., when an agent increases the number of credit cards they have, they have not actually improved their creditworthiness; they have merely tried to game the credit scoring system). This meant that \emph{every} strategic adaptation was perceived as ``gaming'' and hence, the principal was trying to suppress it. However, for some applications of interest (e.g., for school admissions or loan approvals), some types of strategic adaptation are not gaming and should instead be encouraged or incentivized. For example, in a school admissions example, a strategic adaptation that makes the student study more in order to pass the threshold for admission is not gaming; rather, it a way for the student to become a better potential candidate for the school of their choice.

To capture this, some settings in incentive-aware ML assume that in any $d$-dimensional feature vector, some features are \emph{causal} (i.e., by changing them, the agent can change their actual $y$) while the rest are \emph{proxy}/\emph{non-causal} (i.e., by changing them, the agent cannot change their actual $y$). As a result, agent actions that change causal features have the ability to change the ground truth qualifications of an agent; as such, they can lead to \emph{genuine improvement}, as opposed to the \emph{gaming} which is induced by proxy features. The papers that assume causality of features (e.g., \cite{miller2020strategic,shavit2020causal,bechavod2021gaming}) use the language of \emph{structural causal graphs} \cite{pearl2009causality} in order to model the causal effects of the agents' different features.

\subsection*{Agents' Response}

We next turn our attention to the way in which the agents choose their best response to the principal's algorithm. For an agent with ground truth feature vectors $x$, we use $u(x, \hx; f)$ to denote the agent's \emph{utility} for reporting $\hx$ when the principal uses classification/regression function $f$. We focus on utility functions of the form: 
\begin{equation}\label{eq:utility-def}
u(x, \hx; f) := \val (\hx; f) - \cost(x, \hx)
\end{equation}
where $\val(\hx,y;f)$ corresponds to the \emph{value} that the agent obtains by reporting $\hx$ when the principal uses $f$, and $\cost(x, \hx)$ corresponds to the \emph{cost} that agent incurs for changing their feature from $x$ to $\hx$. There have been two types of value functions that have been primarily used in the literature: 
\begin{enumerate}[label=(\roman*)]
\item (continuous) $\val(\hx; f) := f(\hx)$ (i.e., the value is just the evaluation of the function $f$ for the reported feature $\hx$) (e.g., \cite{dong2018strategic,bechavod2022information,shavit2020causal}). 
\item (discrete) $\val(\hx;f) :=  \gamma \cdot \1 \left\{ \sign(f(\hx)) = 1\right\}$ (i.e., the agent cares only about being classified as passing a threshold (e.g., \cite{chen2020learning,ahmadi2021strategic}). Unless specified otherwise, we will use $\gamma = 1$.
\end{enumerate}
As for the cost function, there have been primarily two families that the literature has considered: 
\begin{enumerate}[label=(\roman*)]
\item ($L_p$-norm) $\cost(x, \hx) := \delta \cdot \| x - \hx\|_p$ for some $\delta > 0$ (e.g., \cite{chen2020learning,ahmadi2021strategic,bechavod2021gaming}). The most frequently used norms are $p = 1$ and $p = 2$. 
\item (separable) $\cost(x, \hx):= c(\hx) - c(x)$ (e.g., \cite{hardt2016strategic,hu2019disparate}). These cost functions are suitable for settings where achieving each feature has a certain cost, but this is independent of which feature the agent started from. 
\end{enumerate}

The vast majority of the literature assumes that (given the aforementioned utilities) the agents are \emph{best responding} to $f$, i.e., that $\hx(f) = \arg \max_{x' \in \calX} u(x, x'; f)$. There are some notable exceptions to this assumptions which we highlight in Section~\ref{sec:robustness}. Finally, most of the literature (except a few, e.g., \cite{dong2018strategic}) assumes that \emph{all} agents can respond strategically. For example, in classification, even agents with $y=1$ may want to strategize, if they know that the classification rule $f$ will classify them as $0$.

\begin{remark}\label{remark:principal-knows}
    In general, we assume that the principal knows the agents' value and cost functions (including $\delta$); they are only missing the original point $x$ and can never fully learn it. To be more specific, given the value and cost functions, the reported $\hx_t$ and the $y$, the principal \emph{cannot} reverse engineer the original $x$. There are a couple of works that focus on restricted strategic classification settings where $\delta$ is unknown, but the principal can still learn robust decision rules (see Section~\ref{sec:robustness} for details).
\end{remark}

\subsubsection*{Continuous Adaptation vs Manipulation Graph} Some works move away from the continuous\footnote{The literature sometimes refers to this type of strategizing as ``ball manipulation''.} model of strategic adaptation. Instead, they introduce the idea of a \emph{manipulation graph} (e.g., \cite{ahmadi2023fundamental}). In incentive-aware learning with manipulation graphs, the assumption is that there exists a graph $G(\calX, E)$ to capture all possible manipulations. In graph $G$, each node corresponds to a different feature vector and each edge $e = (x, x') \in E$ captures the manipulation from $x$ to $x'$. The cost function then $\cost(x, x')$ is defined as the sum of costs to move from $x$ to $x'$, if such a path exists in $G$. We will highlight which works use manipulation graphs instead of continuous adaptation in the coming sections.

\subsubsection*{Full vs Partial Information about the Principal's Algorithm} We have so far assumed that the agent has \emph{full} knowledge of $f$ (or $f_t$) at the time of choosing their best response.\footnote{Historically, this is a byproduct of the fact that the original papers modeled the paper as a Stackelberg game. In the Stackelberg games literature, the standard assumption is that the principal announces their strategy at the beginning of the interaction with the agent. This announcement gives them ``commitment power'' (as it is referred to in that literature).} Although this is a useful assumption to understand what solutions are possible in the worst case, in reality it is far from the truth; while agents do exhibit strategic adaptation, they seldom have \emph{full} information about the decision-making rules used. There has been an emerging interest in modeling partial information from the agent side (e.g., \cite{braverman2020randomness,ghalme2021strategic,bechavod2022information,cohen2024bayesian}), but no single model has prevailed as the canonical one. We highlight these models in the coming sections.

\subsection*{Heterogeneous Agents}

Finally, we have so far assumed that there is a single $\calD$ representing the entire population and that every agent shares the same utility function. In other words, we have assumed that agents are \emph{homogeneous}. However, this assumption is often unrealistic; for instance, in the context of school admissions, it is unlikely that everyone in the population has the same natural ability to succeed in school or the same capacity to take steps to improve their chances of being admitted.

Agent heterogeneity has been studied primarily in two different forms. First, agents may come from heterogeneous populations (i.e., their features and labels may originate from different distributions e.g., \cite{milli2019social,hu2019disparate}). Second, agents may have different abilities to adapt to the decision rule that the principal is using (either because of different cost functions e.g., \cite{milli2019social,hu2019disparate}) or because of different understanding of the decision rule (in the case of partial information) (e.g., \cite{bechavod2022information}). We discuss heterogeneous agents in Section~\ref{sec:fairness}.

\begin{remark}
As should be clear by now, this article focuses exclusively on strategic adaptation that occurs in the \emph{feature} space, rather than the \emph{label} or \emph{response variable} space. There have also been a series of works on ML algorithms when the agents can strategically adapt their label (e.g., \cite{dekel2010incentive,chen2018strategyproof}) but they are beyond the scope of this article. The aforementioned articles take a ``robustness'' perspective.
\end{remark}

One final note: the terminology introduced in this section will be used throughout the following sections to describe each paper. This consistent terminology is intended to help the reader develop a clear mental framework for understanding the types of results obtained for each model variant of incentive-aware learning.

\section{Robustness Perspective Main Results}\label{sec:robustness}

We begin our exposition with the \emph{robustness} perspective. In this framework, the principal seeks to learn the most accurate decision-making rule (as defined in Equation~\eqref{eq:accuracy-offline}) that maps agent features to a score or classification label, thereby minimizing their loss. Simultaneously, agents strategically manipulate the data they submit to the decision-making rule in an effort to ``game'' the system. We first examine the offline/batch and online learning settings in Sections~\ref{sec:offline} and~\ref{sec:online}, respectively, focusing on scenarios where agents have full knowledge of the principal's decision-making rule. Subsequently, in Section~\ref{sec:partial-info-principal}, we explore settings where agents have only \emph{partial} information about the principal's decision-making rule. Finally, we conclude this section by discussing cases where agents are \emph{not} individually rational when selecting their misreports, $\hat{x}(f)$, in Section~\ref{sec:beyond-rationality}.

\subsection{Offline and Batch Learning Setting}\label{sec:offline}

\cite{hardt2016strategic} introduced the problem of ``strategic classification'' in the \emph{offline} setting and formulated it as a Stackelberg game. In their framework, the population of agents is assumed to be \emph{homogeneous}, with each agent aiming to maximize their probability of being classified as $+1$ while incurring a cost for doing so. The principal, on the other hand, wants to design a classifier that converges to the offline optimal in terms of ``accuracy'' (as defined in Equation~\eqref{eq:accuracy-offline}) for the $0-1$ loss. The agents are assumed to have \emph{full} information about the classification rule and are best-responding to it. The authors show that for agents with separable cost functions, it is possible to design efficient and nearly optimal classifiers, even for concept classes that are computationally hard to learn. Their theoretical framework further includes impossibility results for learnability when the agents have \emph{general} cost functions, illustrating the fundamental challenges of achieving classification robustness against strategic behavior.

Working in the \emph{offline} or \emph{batch} setting with a \emph{homogeneous} population of agents, \cite{levanon2021strategic} introduce the notion of \emph{strategic empirical risk minimization} (strategic ERM) as an approach for designing strategy-robust decision rules for the principal. At a high level, the authors propose a ``smoothed'' version of the strategic classification problem, incorporating the agents' best-response behavior as a function of the decision rule $f$ into the optimization process for $f$. While the paper does not provide theoretical guarantees, it includes a series of experiments demonstrating how strategic ERM might perform in practice. However, the assumption of a ``smoothed'' version of the problem has limitations from a real-world modeling perspective. As noted by several works (e.g., \cite{dong2018strategic,chen2020learning}), the motivating settings for strategic classification often make it infeasible to identify a ``smooth'' loss function for the principal once the agents' best-response behavior is incorporated.

Still working within the ERM paradigm, in the \emph{offline} learning setting and drawing intuition from traditional PAC learning~\cite{valiant1984theory}, there has also been interest in a PAC version of incentive-aware learning, i.e., given a set of points that have been strategically modified, identify the complexity of finding a classification function that is $\eps (\eta)$-optimal (according to Equation~\eqref{eq:accuracy-offline}) with high probability at least $1 - \eta$. This version of the problem was introduced by \cite{zhang2021incentive}. The authors assume that the agents can best-respond according to a \emph{reporting structure} which maps original features to manipulated ones.\footnote{This can be considered as part of the general ``manipulation graph''-type of cost functions.} Moreover, they assume that the principal is facing a \emph{homogeneous} population of agents. The paper first shows that the vanilla ERM (i.e., the one ignoring incentives) has poor performance in strategic settings.\footnote{For the online setting, a slightly stronger result of two-way incompatibility between regular and strategic settings was obtained by~\cite{chen2020learning}. Specifically, the authors show that there exist classification settings for which every no-external regret algorithm incurs linear Stackelberg regret and vice versa.} Subsequently, they show that a version of \emph{strategic empirical loss} can obtain nearly optimal sample complexity bounds. To construct their strategic empirical loss, the authors take a ``worst-case perspective''; for each reported point, they substitute it with the worst-possible original point it could have originated from.\footnote{A version of this technique was also used in~\cite{chen2020learning}, albeit for the online version of the problem.}

In a similar vein, \cite{lechner2022learning} study the learnability of general concept classes with a new class of loss functions called \emph{strategic loss} (which is used as a proxy hypothesis class for the principal). In their setting, the agents can manipulate according to a manipulation graph. The strategic loss is a discrete loss function which takes a value of $1$ every time that either $f(x) \neq y$ (i.e., incorrect classification) or $f(x) = 0$ but there exists a point $x'$ such that $x'$ is a reachable misreport from $x$ and $f(x') = 1$, and $0$ otherwise. This new loss function aims to not only account for accuracy but also, for the societal burden that is induced when the agents fool the classifier.

\cite{sundaram2023pac} take incentive-aware PAC learnability one step further; the agent population is now \emph{heterogeneous} (i.e., the cost function is the same across agents, but each agent may have a different $\gamma$ in their value function), the principal does \emph{not} know the agents' value functions, but the principal has access to a training dataset that is un-manipulated (i.e., the principal can see some original $x$'s). The key contribution of the work is the introduction of the \emph{Strategic VC-Dimension} (the strategic analogue of VC-dimension), which quantifies learnability in settings where test data is strategically manipulated based on \emph{heterogeneous} agents. The authors subsequently characterize the statistical and computational limits of strategic linear classification. This study also explores the role of \emph{randomization} in improving accuracy under strategic manipulation. We expand on the role of \emph{randomness} in strategic classification settings in Section~\ref{sec:partial-info-principal}.

\cite{rosenfeld2024oneshot} focus on learning a \emph{linear} classifier (the principal has access to a set of un-manipulated data at training time), the agents have a $0-1$ value function, and $L_p$-norm cost function. Importantly, the authors assume that $\delta$ (i.e., the cost function) is \emph{not} known by the principal but is the same\footnote{This is the main difference with the \emph{model} of~\cite{sundaram2023pac}.} across all agents; yet, the principal still needs to learn a classification rule that converges to the optimal one. The authors take a robust optimization approach, by minimizing the worst-case risk over a family of costs which includes the target (unknown) cost. They do so, because as they show, if the principal
has to commit to a single fixed cost for their risk minimization problem, then ERM can never provide a non-trivial data-independent guarantee (unless the assumed single fixed cost were precisely correct). As for the ERM, the authors consider a type of \emph{hinge} loss, that is appropriately expanded in order to include the uncertainty induced by the unknown cost function. The main result of the paper is an efficient iterative algorithm that converges to the minimax optimal solution with rate $\tilde{O}(1/\sqrt{T})$, where $\tilde{O}(\cdot)$ hides polylogarithmic terms, and $T$ is the number of the algorithm's iterations.

\subsection{Online Learning Setting}\label{sec:online}

The online learning version of strategic classification was first studied by \cite{dong2018strategic}. In their paper, the authors provide linear strategic classification algorithms with sublinear Stackelberg regret (see Equation~\eqref{eq:Stackelberg-regret} against a \emph{homogeneous} population of agents with \emph{linear} values (i.e., the agents care about maximizing their distance from the classifier, while being labeled as $+1$ by it). To give an overview of their approach, let $w_t$ be the normal vector corresponding to classifier $f_t$ for each round $t \in [T]$, i.e., $f_t(x) := w_t^\top x$. The main result of the paper is to find the sufficient conditions on the agents' $\hx(w_t)$ such that $\ell(w_t, \hx_t(w_t))$ is \emph{convex} in $w_t$, when $\ell(w_t, \hx_t(w_t))$ is either the hinge or logistic loss. This task boils down to identifying the sufficient conditions on the agents' cost function in order for $\ell(w_t, \hx(w_t))$ to be convex in $w_t$. Convexity is desired, since if $\ell(w_t, \hx_t(w_t))$ is convex in $w_t$, then the principal can apply any off-the-shelf bandit convex optimization algorithm and obtain sublinear Stackelberg regret. The paper obtains improved regret bounds under the assumptions that all agents with $y_t = 1$ are non-strategic. 

But what happens when $\ell(w_t, \hx_t(w_t))$ is not a convex function of $w_t$? In an effort to answer this question in a general way, \cite{chen2020learning} studied online learning of linear classifiers in the following setting: the agents have a discrete value for passing the classifier (i.e., they obtain a value of $1$ for passing the classifier and $0$ otherwise), their cost function is $\delta$-bounded (i.e., $\|\hx_t(f_t) - x_t \| \leq \delta, \forall t \in [T]$), and the learner cares about the $0-1$ loss. Importantly, the results of the paper do not require the agents to \emph{rationally} best-respond; instead, knowing that the $\hx_t(f_t)$ satisfy the constraint that $\|\hx_t(f_t) - x_t\|_2 \leq \delta$ is enough. The paper provides a nearly tight algorithm that dynamically and adaptively partitions the space of feasible classifiers for the principal as new agents arrive. The final Stackelberg regret bound depends on the \emph{instance} of datapoints $\{(x_t, y_t)\}_{t \in [T]}$ that nature chooses. The key trick that the authors use is that when the principal sees a reported point $\hx_t(f_t)$, then they know for sure that the true $x_t$ lies inside a ball $B$, where $B : = \{ x \in \calX:  \| x - \hx_t(f_t) \|_2 \leq \delta \}$. The final trick is to observe that given this information and the fact that the learner cares about the $0-1$ loss, then the principal can obtain perfect information about the loss that would have been incurred in that round $t$ if the same agent at round $t$ were to best respond to some other normal vectors $w$ for which $\| \hx_t (w) - \hx_t(f_t) \|_2 \leq 2 \delta$. The theoretical analysis of the algorithm requires knowing the magnitude of the agents' manipulation ($\delta$) and access to a carefully crafted oracle that can provide some extra information to the principal about the structure of the agents' unmanipulated data.

The aforementioned paper trades efficiency for generality. When the sequence of data $\{(x_t, y_t)\}_{t \in [T]}$ chosen by nature is \emph{separable} by a margin, \cite{ahmadi2021strategic} introduce a variant of the Perceptron algorithm, called the \emph{Strategic Perceptron}, which is \emph{computationally efficient} and converges to a maximum-margin classifier while making a bounded number of mistakes. The upper bound on the number of mistakes depends on the margin of the original, unmanipulated data and the agents' strategizing power. The Strategic Perceptron is analyzed under the assumption that agents incur either $L_1$ or $L_2$ costs when misreporting from $x$ to $\hx$, and are rationally best-responding. Notably, the paper shows how to leverage the structure of the agents' utility function together with the fact that the agents are rationally best responding to establish bounded mistake guarantees \emph{even when} the magnitude of the manipulation cost is \emph{not} known to the principal a priori --- a result that was not achievable in \cite{chen2020learning}.

Next, we transition from models of continuous strategic adaptation to models where agents determine their $\hx_t$ based on a manipulation graph, highlighting the work of \cite{ahmadi2023fundamental}. This setting generalizes the frameworks of \cite{zhang2021incentive} and \cite{lechner2022learning} to the online setting. The paper demonstrates that, unlike in the non-strategic classification setting, the vanilla Halving algorithm may incur an infinite number of mistakes. To address this, the authors propose a general algorithm for the strategic setting with a mistake bound of $O(\Delta \ln (|\mathcal{H}|))$, where $\Delta$ is the degree of the manipulation graph and $\mathcal{H}$ is the (known) class of the target function. Furthermore, the paper extends the algorithm to the agnostic learning setting.

Adopting a similar perspective of testing the limits of strategic learnability, \cite{cohen2024learnability} and \cite{ahmadi2021strategic} investigate whether the learnability of a concept class implies its strategic learnability. They essentially show that every learnable function class remains learnable even when data is strategically manipulated. Both works model the agents' feasible manipulations using manipulation graphs and consider scenarios where the graph is either fully known or only partially known to the principal. \cite{ahmadi2021strategic} introduce the ``strategic Littlestone dimension,'' which captures the complexity of the agents' manipulation graph and the hypothesis class. Both papers analyze strategic learnability across multiple variations of the baseline strategic classification model. Finally, \cite{shao2024strategic} study learnability in terms of mistake bounds and sample complexity when agents' manipulations are \emph{heterogeneous}. They consider both continuous adaptations and manipulation graphs. As for the principal, they assume that some knowledge of $x_t$ is available either before choosing the classification rule $f_t$ or immediately afterward.

In a slightly different setup, \cite{harris2023strategic} consider an online setting where at each round the principal commits to a function $f_t: \calX \to \{0,1\}$, the agents can strategically adapt within a ball of radius $\delta$ of their true datapoint $x_t$, and the reward that the principal receives is linear in the agent's unmodified context; more concretely, for each decision $\alpha \in \{0,1\}$ the reward of the principal for a context $x_t$ is: $r_t(\alpha) = \theta_{\alpha}^\top x_t + \eps$, where $\theta_\alpha$ is a $d-$dimensional vector. The authors assume that the principal has ``apple tasting'' feedback, i.e., the principal can observe $r_t(\alpha)$ only when $\alpha = 1$ (which in turn, is decided by the function $f_t$). The authors present algorithms that actually incentivize agents to be \emph{truthful} (i.e., report $x_t$ without any manipulation) while achieving sublinear regret.

\subsection{Partial Information about the Principal's Algorithm}\label{sec:partial-info-principal}

So far, we have primarily assumed that the principal commits to a \emph{deterministic} rule and that agents fully observe this rule. \cite{braverman2020randomness} were the first to highlight the role of \emph{randomness} in the principal's classifier and the impact of \emph{noise} in the agents' features on the outcomes of the strategic classification game. The paper demonstrates that to maximize accuracy (as defined by Equation~\eqref{eq:accuracy-offline}), the principal may \emph{need} to employ randomized rules. This result creates an intriguing policy dilemma: on the one hand, randomized rules may be necessary to achieve optimal accuracy; on the other hand, their deployment can be legally problematic. Interestingly, the paper shows that introducing (or having inherently) noisier signals for the agents' features can improve both accuracy and fairness in equilibrium across different subpopulations. 

\cite{ahmadi2023fundamental} also explore the role of randomness in strategic classification, focusing on its impact on learnability. They consider two sources of randomness. In the first, the principal commits to a probability distribution over classifiers, thereby inducing certain probabilities of classification as $+1$ for agents. In the second, the principal commits to a probability distribution over classifiers, nature (which may adversarially select the next $x_t$) responds to this distribution, and the chosen agent $x_t$ best responds to the \emph{realized} classifier. The second model is more \emph{transparent} to the agents than the first and enables the principal to design algorithms with improved regret guarantees.

If the principal has the choice between a transparent and an ``opaque'' classifier, which approach minimizes prediction error? \cite{ghalme2021strategic} address this question in the setting of \cite{hardt2016strategic} (i.e., offline, homogeneous population of agents, etc.). They define the \emph{price of opacity} as the difference in prediction error when agents respond to a fully transparent classifier $f$ versus an opaque rule $\hat{f}$. The paper studies the conditions under which the price of opacity can be positive or negative. Consistent with the theory of Stackelberg games, revealing $f$ (or allowing it to be fully anticipated or deduced from $\hat{f}$) can sometimes benefit the principal, as it enables them to precisely predict how agents will react.

\cite{cohen2024bayesian} introduce a Bayesian classification setting, where the principal gradually reveals information about the classification rule. In this model, agents share a common distributional prior over the classifier used by the principal and best respond by maximizing their expected utility. The principal, in turn, can strategically release partial information about the classifier over time. The authors show how to release this information carefully to ensure that truly qualified agents (i.e., $y_t = +1$) can pass the classifier while preventing unqualified agents from gaining sufficient information to successfully strategize and game the system.

Finally, \cite{bechavod2022information} study a setting where agents acquire information about the classifier through ``peer learning.'' The primary focus of this work is on the fairness implications of information discrepancies across different subpopulations. Therefore, we defer a detailed discussion of this work to Section~\ref{sec:fairness}.

\subsection{Beyond Rational Best-Response Agents}\label{sec:beyond-rationality}

So far, we have focused on settings where agents best-respond to the principal's rule. We now shift our attention to scenarios where agents do \emph{not} precisely best-respond.

Although the results in \cite{chen2020learning} hold for \emph{any} agent manipulation within $\delta$ of the true data point, \cite{jagadeesan2021alternative} formalize alternative models for agent behavior that deviate from exact, rational best response. The authors demonstrate the brittleness of standard strategic classification algorithms when agents do not strictly adhere to the assumed best-response model. To address this, they identify a set of desiderata for agent responses that ensure algorithm stability and propose the \emph{noisy response} model. In this model, agents best respond to a noise-perturbed version of the decision rule, inspired by the principles of smoothed analysis~\cite{spielman2009smoothed}.

\cite{ebrahimi2024double} study the role of behavioral biases in agents' responses within strategic classification settings. Specifically, they consider agents who, when evaluating the value of passing the classifier, \emph{weigh} the classifier's features according to their own biases. The paper analyzes a homogeneous population of agents who can incur a cost of up to $B$ for misreporting. It identifies cases where agents overshoot or undershoot the classifier's boundary due to their biased perceptions of the classifier's feature weights.

\cite{lechner2023strategic} examine settings where the principal faces two sources of uncertainty regarding the agents' responses. First, agents are not required to rationally best-respond and are instead permitted to use any \emph{feasible} response that enables them to fool the classifier. Second, the principal does not have full knowledge of the agents' manipulation graph but only knows the general family to which it belongs. Focusing on strategic loss, the authors study the learnability of both proper and improper learning under these assumptions. Their key result is that it is possible to learn an almost-optimal classifier in terms of strategic loss, even without precise knowledge of the manipulation graph.

\cite{cohen2024learnability} explore the effects of partial knowledge of the manipulation graph on learnability. They show that when the principal knows only the general family of graphs to which the manipulation graph belongs, they can achieve nearly tight bounds on both sample complexity and regret. Furthermore, the difference in learning complexity between the fully-known and partially-known graph settings is (roughly) logarithmic in the size of the graph family.

Finally, \cite{ahmadi2024strategic} also assume that the principal knows only the \emph{family} of graphs to which the agents' manipulation graph belongs. They derive a regret bound that is approximately optimal for certain instances. Additionally, they extend their results to a setting where each agent may have a different manipulation graph, provided all graphs belong to the same family. This generalized setting is referred to as the ``agnostic'' case.

\section{Improvement \& Causality Perspective Main Results}\label{sec:causality-improvement}

Oftentimes, strategic adaptation to algorithmic decision-making rules may lead to genuine improvement for the individuals; for instance, paying-off prior debt as a means of increasing your credit score actually helps you become more creditworthy. To state this in the language of incentive-aware learning, this means that the agents' label $y$ can change when they switch from their true $x$ to the strategically manipulated $\hat{x}$. This section focuses on settings where strategic adaptation can lead to both gaming and actual improvement.

\subsection{Improvement \& Recourse}\label{sec:improvement-recourse}

According to the ``improvement''/``recourse''\footnote{The term ``recourse'' comes from the traditional ML literature. Loosely speaking, algorithmic recourse refers to the ability of individuals to reverse negative decisions made by algorithms through counterfactual explanations provided alongside the decision. A substantial body of work exists on algorithmic recourse (see, e.g., \cite{karimi2020survey} for a survey), but it is beyond the scope of this article. Here, we focus specifically on the effects of strategic adaptation on algorithmic recourse.} perspective, any strategic adaptation results in genuine improvement for individuals; that is, when a data point changes from $x$ to $\hx$, it holds that $\hst(x) < \hst(\hx)$.

\cite{kleinberg2020classifiers} introduce one such model where agent ``manipulations'' result in changes to the underlying features, which can constitute genuine improvement for the agents. Their primary result shows that simple linear mechanisms suffice to incentivize genuine improvement in settings where the principal interacts with a single agent. \cite{harris2021stateful} extend the model of \cite{kleinberg2020classifiers} to settings where agents achieve improvements over a sequence of rounds, i.e., agents transition through different states over time as they respond to the principal's rule.

\cite{alon2020multiagent} generalize the single-agent setting of \cite{kleinberg2020classifiers} to a multi-agent framework. In their model, all agents share the same initial feature representation, but their ability to manipulate (quantified by their manipulation costs) differs.

\cite{haghtalab2020maximizing} also study multi-agent settings, focusing on designing evaluation mechanisms that maximize population-wide quality scores when agents can strategically alter their features at a cost. Their model differs from \cite{alon2020multiagent} in that agents can have different initial feature representations. The authors analyze two specific classes of mechanisms: linear mechanisms and linear threshold mechanisms. For linear mechanisms, they show that the optimal strategy corresponds to projecting the true quality function onto the observable feature space, which is computationally efficient. For linear threshold mechanisms, they develop approximation algorithms, including a constant-factor approximation algorithm under smooth feature distributions, that balance the trade-offs between incentivizing improvements and maximizing welfare. The paper further considers scenarios where the feature distribution is unknown and provides sample-complexity guarantees for learning optimal mechanisms.

\cite{tsirtsis2020decisions} explore the design of optimal decision-making policies and counterfactual explanations in incentive-aware learning. They model this problem as a Stackelberg game, where decision-makers provide \emph{counterfactual explanations}—guidelines on how agents can change their features—and agents respond strategically to maximize their benefit. Unlike the standard Stackelberg game for incentive-aware learning, where the decision rule is announced, here the principal announces counterfactual explanations. The authors show that optimizing the set of counterfactual explanations for a fixed decision policy is NP-hard but can be addressed using approximation algorithms that leverage the problem's submodularity. They further extend the problem to jointly optimize both the decision policy and explanations, reducing it to a non-monotone submodular maximization problem solvable with approximation guarantees. Additionally, the paper incorporates diversity constraints to ensure equitable distribution of explanations across populations.

Finally, \cite{bechavod2022information} study the ``improvement'' perspective when the principal's decision rule is not fully known to the agents. Their work focuses on the effects of information discrepancies across different subpopulations and is therefore discussed in Section~\ref{sec:fairness}.

\subsection{Causality}

How can we distinguish between agent actions that lead to genuine improvement versus those that constitute mere gaming? As \cite{miller2020strategic} observe, designing ``good'' incentive-aware decision-making rules—rules that incentivize actions leading to genuine improvement while disincentivizing gaming—is equivalent to identifying the causal model underlying the setting (i.e., performing causal inference). Their work was the first to formalize this connection, introducing causal graphs to study how certain agent features affect (or do not affect) the target variable $y$.

Building on the theme of causality in incentive-aware learning, \cite{shavit2020causal} study incentive-aware linear regression, where the decision-maker seeks to optimize one of three objectives: (1) Agent Outcome Maximization (incentivizing agents to improve their outcomes), (2) Prediction Risk Minimization (ensuring accurate prediction of post-gaming outcomes), and (3) Parameter Estimation (accurately estimating the causal parameters of the outcome-generating process). The authors propose efficient algorithms for each objective in a realizable linear setting, leveraging the ability to test and observe agent responses to decision rules—effectively performing causal interventions. This ability to perform interventions makes their setting more tractable compared to \cite{miller2020strategic}. Additionally, they address challenges such as omitted variable bias and interactions between observed and hidden features, which can undermine naive regression approaches. Extending beyond linear regression, \cite{horowitz2023causal} study (agnostic) incentive-aware classification under causality with the goal of improving the principal's accuracy.

In concurrent and independent work, \cite{bechavod2021gaming} explore incentive-aware linear regression and demonstrate how agents' strategic behavior can facilitate the learning of causal variables. The authors propose a batch-based retraining approach that iteratively updates the regression model, leveraging agents' strategic modifications to improve predictive accuracy while incentivizing genuine improvement in features. They prove that this dynamic interaction enables the principal to accurately recover the true regression parameters over time, even when features are correlated. As a result, the principal can both incentivize genuine improvement and improve the robustness of the decision model.

Finally, \cite{ahmadi2022class} study incentive-aware classification under a causal model, addressing both discrete strategic adaptation (via manipulation graphs) and continuous adaptation. For the general discrete model, the authors design efficient algorithms to maximize true positives while ensuring no false positives, thus guaranteeing that only genuinely qualified agents are classified positively. They further show that the problem of selecting criteria to maximize true positives while allowing even a bounded number of false positives becomes NP-hard. In the continuous adaptation (linear) model, they develop algorithms to determine whether a linear classifier exists that classifies all agents accurately while incentivizing all improvable agents to become qualified.

\subsection{Performativity}

Before we conclude the section on improvement and causality in incentive-aware ML settings, we briefly touch on the literature on \emph{performative prediction} \cite{perdomo2020performative}. Performative prediction is another framework to explain and reason about how predictions, when used to inform decisions, influence the outcomes they aim to predict. The authors develop a risk minimization framework and propose a new equilibrium notion called performative stability. Roughly speaking, this notion ensures that predictions are calibrated not to past data but to the outcomes they induce. The paper presents necessary and sufficient conditions for retraining algorithms to converge to performatively stable solutions with near-minimal loss. The main distinction between performative prediction and the other models that we highlight in this survey is that performative prediction uses certain smoothness assumptions on the way that original points $x$ leads to strategically adapted points $\hx$, instead of focusing on the agents' utility functions.

\section{Fairness Main Results}\label{sec:fairness}

Most (if not all) of the papers discussed so far in this article have focused on a homogeneous population of agents with which the principal is interacting. However, when the principal is interacting with a \emph{heterogeneous} population of agents, with (potentially) different abilities to strategize and different qualifications, then optimizing for the desiderata of robustness-to-gaming or accuracy may have disparate downstream effects to the different subpopulations. 

\cite{hu2019disparate} and \cite{milli2019social} independently and concurrently initiated the study of the disparate downstream effects of designing robust-to-gaming classifiers to different subpopulations. \cite{milli2019social} defined the \emph{social burden} of a classifier as the aggregate of the minimum cost an individual needs in order to be classified as a $+1$. For example, for agents with $y_t = +1$, a high social burden means that it is very costly for the agents to obtain their correct classification. The authors prove a general trade-off between principal's accuracy and agent utility. They also prove that when agents incur cost as a consequence of a principal making their classifier robust to strategic behavior, the costs can disproportionally fall on the disadvantaged subpopulations. 

In a similar theme, \cite{hu2019disparate} study negative externalities of strategic classification, and show that the Stackelberg equilibrium classifier leads to only false negative errors on the disadvantaged subpopulation but only false positives on the advantaged population. Not only that, but they also show that providing a cost subsidy (whose goal is to counterbalance this the difference in false negatives and false positives from each subpopulation) can \emph{actually} lead to worse outcomes for \emph{everyone} in the game.

Focusing on the goal of group fairness, \cite{estornell2023group} explores the unintended consequences of using fairness-aware algorithms in environments where agents can strategically manipulate their features to achieve better outcomes. While fairness in algorithmic decision-making is typically aimed at ensuring equitable treatment across demographic groups, the paper identifies a phenomenon called ``fairness reversal''. This occurs when a fairness-driven classifier (designed to equalize outcomes between groups) becomes less fair than a conventional accuracy-focused classifier due to strategic feature manipulation by agents. The authors empirically demonstrate this phenomenon using benchmark datasets and attribute it to the selectivity of fair classifiers, which achieve fairness by excluding individuals from the advantaged group rather than including more from the disadvantaged group. They prove that increased selectivity is a sufficient, and in some cases necessary, condition for fairness reversal. They further show that fairness reversal does not occur when fairness is achieved through inclusiveness, where the classifier broadens access to the disadvantaged group.

The focus of the aforementioned works was on fairness in terms of classification accuracy. Lately, some works have started considering fairness in terms of improvement or recourse ability. \cite{gupta2019equalizing} address fairness in terms of \emph{recourse}, i.e., the effort required to reverse a negative classification, across demographic groups. Mathematically, recourse is measured as the distance from an individual's features to the decision boundary of a classifier. The paper introduces a new approach to regularize classifiers, minimizing disparities in recourse while maintaining predictive accuracy. It extends prior work on linear classifiers to more complex settings, including non-linear models and model-agnostic scenarios, where the decision boundary is not explicitly known. For the model-agnostic setting, the paper assumes that the agents have black-box access to the classifier, rather than the full mathematical formulation.

\cite{bechavod2022information} study how disparities in information about decision rules affect the ability of agents from different sub-populations to improve their outcomes in strategic learning contexts. Unlike most traditional models that assume agents fully know decision rules, this work focuses on scenarios where decision rules are not fully known originally, and agents infer them based on their peers' experiences, creating group-specific information levels; they refer to this process as ``peer learning''. The study reveals that even when decision rules are optimized to maximize welfare, disparities in information and effort costs can lead to some sub-populations experiencing a decline in quality (``negative externality''). However, under specific conditions (e.g., proportional costs across groups or minimal information overlap --- measured through an ``information overlap proxy'' metric --- across groups) negative impacts can be mitigated. 

\cite{ahmadi2023fairincentives} study the problem of designing short-term goal structures to incentivize agents with varying abilities to improve their skills or capacities-for-improvement. It proposes two models: (1) the common improvement capacity model, where all agents share the same improvement limit, and (2) the individualized improvement capacity model, where agents have personalized improvement limits. The authors develop algorithms to optimize the placement of target skill levels (i.e., goals) to maximize social welfare (i.e., total improvement across all agents) and ensure fairness among groups. One challenge they address is the non-monotonic nature of social welfare, where adding new target levels may unintentionally reduce overall improvement. Finally, they present an extension for the case where the principal has sample access to the available data when designing the classifier.

\section{Conclusion}\label{sec:conclusion}

The purpose of this article has been to provide a gentle introduction to the exciting area of incentive-aware ML. We categorized the existing research into \emph{robustness, fairness}, and \emph{improvement/causality perspectives}, and we highlighted the diverse approaches and objectives within each domain. We outlined some of the foundational models and theoretical frameworks for understanding strategic adaptation, from offline and online learning settings to causal perspectives, and we emphasized the complexities introduced by agent heterogeneity and partial information.

There have also been a handful of topics related to incentive-aware ML settings that we did not touch upon, as they did not directly fit under one of our three outlined perspectives. Examples include: \cite{zrnic2021leads} who study how the Stackelberg game (and its outcomes) change when the principal and the agent (termed ``leader'' and ``follower'' in their paper) alternate in order; papers on econometrics for strategic agents (e.g., \cite{harris2022strategic,harris2024strategyproof}); and papers focusing on agents that can choose to not participate in the algorithmic decision making process, if that is aligned with their utility maximization (e.g., \cite{krishnaswamy2021classification},~\cite{horowitz2024classification}).

For all the excitement surrounding this research area, there is one question that seems as pressing as ever. 

\begin{center}
\emph{What comes next for the literature on incentive-aware ML?}
\end{center}

In the author's view, there are two primary paths for the future of incentive-aware ML. The first path is the more well-established and widely explored. There remain myriad settings requiring theoretical analysis of the interactions between individuals and a decision-making principal. For example, how do information discrepancies about the principal's algorithm across different subpopulations affect their abilities to genuinely improve their outcomes versus merely game the system? Are there properties of ``interpretable'' decision-making algorithms that can provably incentivize genuine improvement rather than gaming? Developing new models and providing provable guarantees for these questions will help solidify the theoretical foundations of incentive-aware ML.

The second path is less charted and relatively unexplored, particularly from a theorist's perspective. Although examples of individuals strategizing and adapting to algorithmic decision-making rules are abundant, incentive-aware ML still needs to identify a \emph{concrete} application domain where the insights gained from theoretical advancements can \emph{actually be applied}. Such a domain would allow incentive-aware algorithms to be deployed and evaluated against other ``robust'' algorithms. 

This approach differs from the path the literature has predominantly taken. To illustrate this distinction, consider the steps required to apply theoretical insights from incentive-aware ML to a practical domain, such as recommendation systems (RecSys).\footnote{Another promising application domain is the health insurance industry, as recently discussed in \cite{chang2024s}.} To apply these insights effectively in the RecSys domain, we would need to address several questions: Do users ``strategize'' with their data (see e.g., \cite{haupt2023recommending})? What utility function are they optimizing for? What does it mean for users to have ``partial'' information about the RecSys? What specific interventions can the RecSys implement to mitigate inequalities between different user subpopulations?

Identifying such a concrete application domain would enable the foundational results in this field to be translated into actionable insights, driving meaningful, real-world change. The author is optimistic about the potential of the next generation of incentive-aware ML research to bridge this gap and create significant societal impact.

\bibliographystyle{plainnat}
\bibliography{strat-class}

\newpage
\appendix

\end{document}